\documentclass[%
 reprint,
 amsmath,amssymb,
 aps,
]{revtex4-2}

\usepackage{graphicx}
\usepackage{dcolumn}
\usepackage{bm}

\usepackage{tikz}
\usetikzlibrary{arrows, positioning, calc}

\usepackage{mathtools}
\usepackage{xcolor}
\newcommand{\tV}{\widetilde{V}}
\newcommand{\CV}{\mathcal{V}}
\newcommand{\CN}{\mathcal{N}}

\definecolor{forestgreen}{rgb}{0.13, 0.55, 0.13}

\begin{document}

\preprint{APS/123-QED}

\title{Stochastic Reconstruction of Gappy Lagrangian Turbulent  Signals \\by Conditional Diffusion Models}

\author{Tianyi Li$^1$}
\author{Luca Biferale$^1$}
\author{Fabio Bonaccorso$^1$}
\author{Michele Buzzicotti$^1$}
\author{Luca Centurioni$^2$}
\affiliation{$^1$Department of Physics and INFN, University of Rome `Tor Vergata', Via della Ricerca Scientifica 1, 00133 Rome, Italy}
\affiliation{$^2$Lagrangian Drifter Laboratory, Scripps Institution of Oceanography, La Jolla, California, USA}



%

\date{\today}

\begin{abstract}
We present a stochastic method for reconstructing missing spatial and velocity data along the trajectories of small objects passively advected by turbulent flows with a wide range of temporal or spatial scales, such as small balloons in the atmosphere or drifters in the ocean. Our approach makes use of conditional generative diffusion models, a recently proposed data-driven machine learning technique. We solve the problem for two paradigmatic open problems, the case of 3D tracers in homogeneous and isotropic turbulence, and 2D trajectories from the NOAA-funded Global Drifter Program. 
We show that for both cases, our method is able to reconstruct velocity signals retaining non-trivial scale-by-scale properties that are highly non-Gaussian and intermittent. A key feature of our method is its flexibility in dealing with the location and shape of data gaps, as well as its ability to naturally exploit correlations between different components, leading to superior accuracy, with respect to Gaussian process regressions, for both pointwise reconstruction and statistical expressivity. Our method shows promising applications also to a wide range of other Lagrangian problems, including multi-particle dispersion in turbulence, dynamics of charged particles in astrophysics and plasma physics, and pedestrian dynamics.
\end{abstract}

\maketitle



%
%

\begin{figure*}[htbp]
	\includegraphics[width=\linewidth]{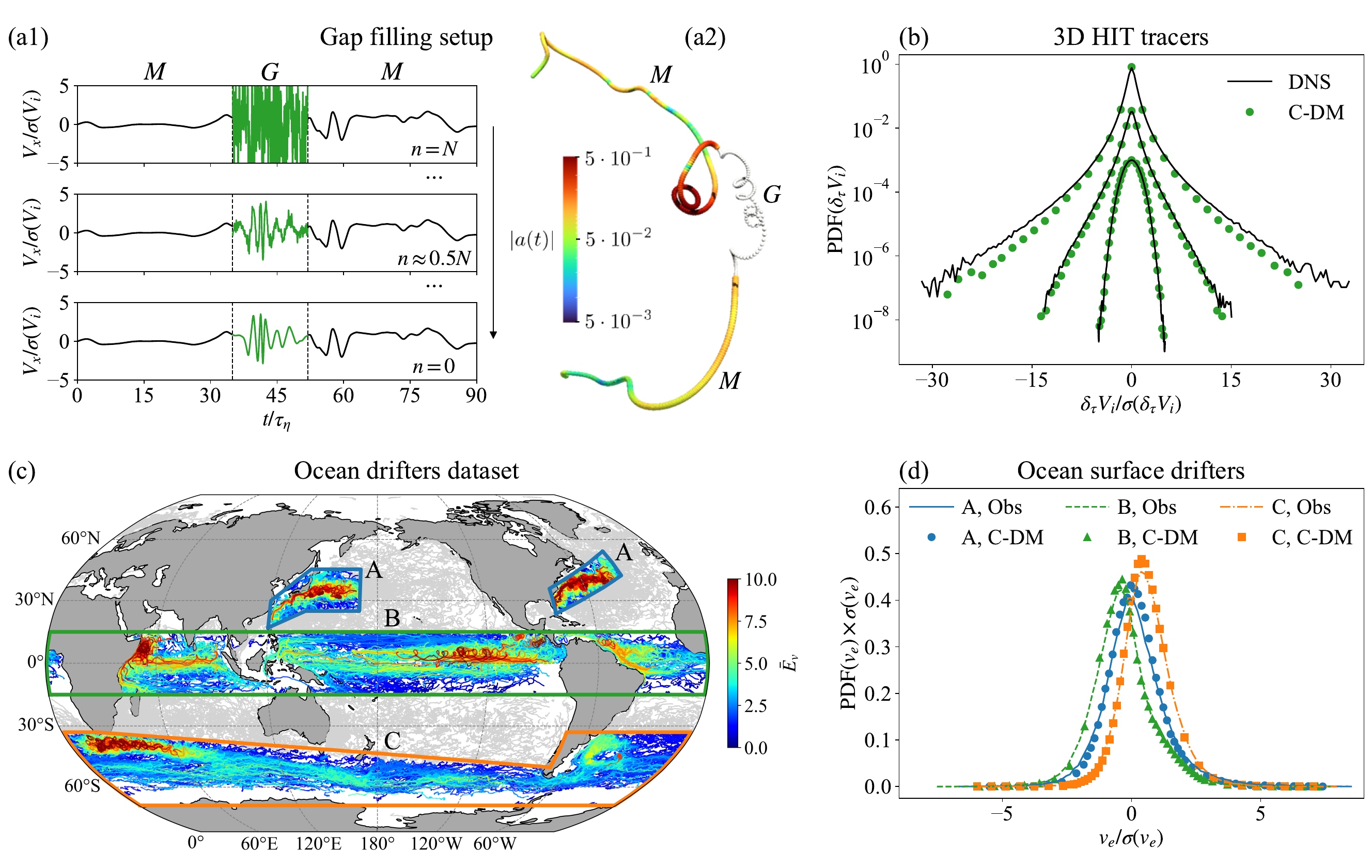}
	\caption{\label{fig:setup}(a1) Setup of the Lagrangian turbulent signal reconstruction. In this example, the goal is to reconstruct the missing observation of one generic velocity component, $V_i(t)$, of a 3D turbulent tracer. We assume that there is missing data within a large time window in the middle of the trajectory (region denoted as $G$), while the beginning and end chunks are assumed to be measured and known (regions denoted as $M$). Once trained, our conditional diffusion model (C-DM) reconstructs the signal within the gap through a {\it backward} multi-step denoising process, starting from a pure uncorrelated Gaussian guess in the region $G$ at the beginning of the process $n=N$ (top row), gradually generating a denoised signal conditioned on the data measure in the regions $M$ (middle row), and ending with the final realistic guess at the last iteration, $n=0$ (bottom row). Panel (a2) shows a 3D representation of the gappy trajectory for visualisation purposes. (b) Standardized probability density functions (PDFs) of a generic component of the velocity increment, $\delta_\tau V_i$, defined in Eq.(\ref{eq:deltav}), for different time lags $\tau/\tau_\eta=0.5,2,100$ (from bottom to top) for both ground-truth DNS data (black lines) and reconstructed data from the C-DM (green solid circles) for a central gap of size $50\tau_\eta$. PDFs for different $\tau$ are shifted vertically for clarity. The PDF is Gaussian for large time lags and develops progressively fatter tails as $\tau$ decreases, illustrating the non-trivial intermittent statistical properties of the Lagrangian turbulence dataset. (c) Ocean surface drifter trajectories \cite{elipot2022hourly}, with three specific regions where trajectories are colored by their total kinetic energy $\bar{E}_v$: (A) two Western Boundary Currents (WBCs), the Kuroshio Current and Gulf Stream (blue contours); (B) the Tropics (TRO) (green contour); and (C) the Antarctic Circumpolar Current (ACC) (orange contour). Trajectories outside these regions are shown in gray. (d) Standardized PDFs of the eastward velocity for the three regions from panel (c), based on observations and C-DM reconstructions, with a central gap of size $360\tau_0$. Observational data (Obs) are shown as blue solid, green dashed, and orange dash-dotted lines, while C-DM reconstructions are shown as blue circles, green triangles, and orange squares for regions A, B and C, respectively. Here, $\sigma$ represents the standard deviation computed from the ground-truth dataset.}
\end{figure*}

Turbulent signals along the Lagrangian trajectories of passively advected objects result from the evolution of complex, multiscale nonlinear interactions involving many excited degrees of freedom \cite{shraiman2000scalar, la2001fluid, falkovich2001particles, yeung2002lagrangian, falkovich2006lessons, toschi2009lagrangian, pomeau2016long, bentkamp2019persistent}. These signals are critical for understanding numerous phenomena across various fields, including geophysical dynamics, combustion, industrial mixing, pollutant dispersion, cloud formation, and cosmic ray propagation \cite{pedlosky1987geophysical, warnatz2006combustion, pope1994lagrangian, xia2013lagrangian, shaw2003particle, zweibel2013microphysics, schlickeiser2015cosmic}. Advection by turbulent three-dimensional (3D) flows is characterised by the presence of wild fluctuations on a large range of time scales, from the largest, $\tau_L$, where energy is injected, to the smallest, $\tau_\eta$, associated with viscous effects. Dynamics at intermediate scales are dominated by nonlinear interactions, with anomalous departures from Gaussianity that become increasingly significant at higher and higher frequencies (see Fig.\ref{fig:setup}b). For quasi-two-dimensional (quasi-2D) geophysical applications, the presence of large-scale coherent structures makes the Lagrangian problem even more complex, with strong influences from boundary conditions and seasonal environmental background \cite{rhines1975waves, mcwilliams1984emergence, pedlosky1987geophysical}. (see Fig.\ref{fig:setup}c,d).

The aforementioned challenges --namely the large embedding dimensions of the emerging dynamics and the non-trivial statistical properties across scales-- make inferring missing information about the Lagrangian properties particularly challenging, especially concerning the predictability of extreme intense events and coherent structures that characterise the intermittent turbulent fluctuations (see Fig.\ref{fig:setup}a). 
Note that the reconstruction problem conditioned on the observed data is typically not unique, as the observed data can be compatible with many possible realisations of the signal within the gap. This is especially true for reconstructing turbulent signals, which reside in a high-dimensional embedding space and exhibit chaotic dynamics and spontaneous stochasticity \cite{bernard1998slow, vanden2000generalized, thalabard2020butterfly, bandak2024spontaneous}. 
For example, oceanic drifters often result in incomplete or `gappy' measurements due to observational constraints and/or communication failures \cite{hansen1996quality, elipot2016global, elipot2022dataset}. Atmospheric turbulence measurements are often sparse and limited to specific spatial points in the wind fields \cite{friedrich2022superstatistical}. Similar challenges exist in areas such as animal movement tracking \cite{wikelski2007going, kranstauber2012dynamic}, pedestrian trajectory prediction \cite{korbmacher2022review}, and cosmic ray propagation in turbulent magnetic fields \cite{zweibel2013microphysics, schlickeiser2015cosmic, friedrich2020stochastic}, as well as in many laboratory setups \cite{adrian1991particle, mordant2001measurement, toschi2009lagrangian}.

Common stochastic reconstruction methods such as kriging \cite{cressie1990origins, oliver1990kriging} and Gaussian process regression (GPR) \cite{williams2006gaussian, foreman2017fast} are based on the knowledge of the covariance matrix and therefore they are optimal only for quasi-normal and self-similar distributions. Similarly, proper orthogonal decomposition (POD) is mainly focused on capturing the properties of energy-containing scales, resulting in a loss of accuracy for small-scale extreme fluctuations \cite{everson1995karhunen, boree2003extended, li2023generative, li2023multi}. To address the multi-scale nature of turbulence, generation and interpolation methods based on fractional Brownian motion and superstatistics with multivariate Gaussian mixture have been proposed \cite{friedrich2020stochastic,beck2003superstatistics,lubke2023stochastic} and shown to capture some of the properties possessed by the original turbulent signals, including multifractality. However, generation/reconstruction methods based on empirical distributions, such as multifractal processes \cite{benzi1984multifractal, meneveau1987simple, biferale1998mimicking, biferale2004multifractal, arneodo2008universal, chevillard2012phenomenological, sinhuber2021multi}, often suffer from epistemic errors and a lack of expressivity, restricting the problem to the case of power-law scaling and failing to optimize the multi-objective physics over the full range of dynamical time scales. As a result, we do not yet have a generic stochastic approach that is flexible and accurate enough to be applicable to reconstruct missing information for Eulerian and Lagrangian turbulent signals.

Very recently, a notable success has been achieved for the {\it unconditional} generation of synthetic Lagrangian turbulence using stochastic data-driven machine learning based on state-of-the-art diffusion models (DMs) \cite{li2024synthetic}. 
These models have demonstrated the ability to reproduce most statistical benchmarks and exhibit strong generalisability for extreme events, including accurate multi-scale properties even beyond the restricted range where pure power laws are observed. They show superiority over other empirical models and the capacity to be easily extended to a variety of different physical applications, such as the trajectories of particles with different inertia \cite{li2024generative}. \\
Here we build on these results and show that it is possible to further extend the applicability of data-driven generative models for Lagrangian turbulence by presenting a stochastic reconstruction method of {\it gappy} signals based on a conditional DM (C-DM). The approach supplements the basic architecture used for unconditional generation with an additional channel that embeds the observed data, enabling the model to {\it stochastically} refill gaps in the original data with the correct correlations (for the C-DM architecture see Fig.\ref{fig:CDM}b in sec.\ref{subsec:C-DM} of the Methods).

Our model provides reconstructions with accurate multiscale statistics from the largest `gappy' scale down to the regime where inertial and dissipative effects overlap, and shows pointwise accuracy for each time inside the gap superior to GPR, especially for the simultaneous reconstruction of all velocity components. We also briefly discuss results for different gap positions (whether in the center of the signal or near its boundaries) and for different gap shapes, including the case of interpolation where the observed data are sampled at a single given frequency (see Supplementary Fig.1). \\
\\
{\sc {\bf Conditional Diffusion Models (C-DMs).}}
DMs, both unconditional and conditional, have recently gained popularity in various fields such as computer vision for image generation and enhancement \cite{dhariwal2021diffusion}, audio generation \cite{oord2016wavenet}, text-to-video synthesis \cite{videoworldsimulators2024}, and have also shown promising results in scientific applications such as bioinformatics \cite{guo2024diffusion}, molecular linker design \cite{igashov2024equivariant}, and quantum circuit synthesis \cite{furrutter2024quantum}, especially in the context of C-DMs \cite{li2023diffusion, hu2024generative, gao2024generative}. \\
To describe our application of C-DMs to refill partially observed Lagrangian velocity signals, we introduce the following notation: each trajectory is defined as $\CV=\{V_i(t_k)|\,t_k\in[0,T]\}$, where $i$ denotes one of the velocity components, and $k=1,\dots,K$ are the discretized sampling times. For each trajectory, the total set of time points is further split into two disjoint sets: $\CV_m=\{V_i(t_m)|\,t_m\in M\}$ and $\CV_g=\{V_i(t_g)|\,t_g\in G\}$, where $M$ and $G$ respectively represent the sets of measured and missing (gap) points, such that $M\cup G=[0,T]$ and $\CV=\CV_m\cup\CV_g$ (see Fig.\ref{fig:setup}a).

The way to proceed is to supplement DM architectures used for generative AI \cite{li2024synthetic} with a conditional framework to ensure that the sampled probability distribution is correctly targeted to match the measured data outside the gap. Specifically, the C-DMs must learn to model the ground truth distribution, $p(\CV_g|\CV_m)$, of $\CV_g$, conditioned on $\CV_m$, such that $p_\theta (\CV_g|\CV_m) \sim p(\CV_g|\CV_m)$, where with $\theta$ we define the set of trained parameters in the C-DM (see sec.\ref{subsec:C-DM} in the Methods). C-DMs consist of a {\it forward} and {\it backward} process. On the one hand, the {\it forward} diffusion process is required to prepare the training dataset and works through an $N$-step Markov chain which gradually adds Gaussian noise to the ground truth signals in the gap (supposed to be available in the training data) until the signal in the gap is reduced to pure Gaussian noise ~\cite{ho2020denoising, saharia2022image, saharia2022palette}. \\

\begin{figure*}[htbp]
	\includegraphics[width=\linewidth]{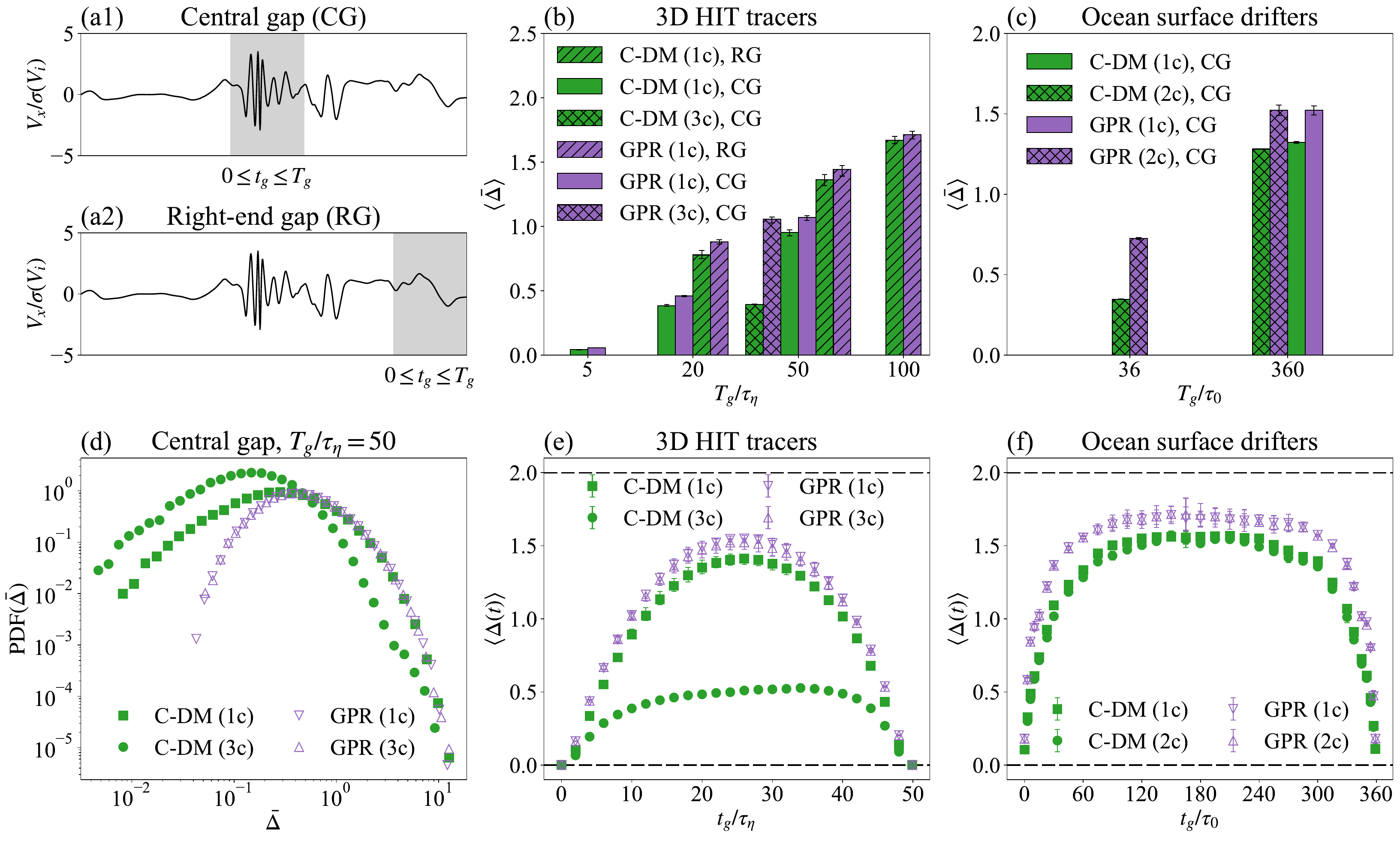}
	\caption{\label{fig:PointRecons} Geometries of (a1) a central gap (CG) and (a2) a right-end gap (RG), with gap regions indicated in gray. (b) Plot of the overall mean squared error (MSE), $\langle\bar{\Delta}\rangle$, for different gaps of sizes $T_g$ for 3D Lagrangian turbulence reconstruction. Results are shown for one generic component (1c) using C-DM (green bars) and Gaussian process regression (GPR, purple bars). Right-end gaps are shown with diagonal hatching, while central gaps are shown without hatching. In addition, for a central gap of size $50\tau_\eta$, the result for 3-components (3c) are also shown, with cross-hatching for C-DM (green) and GPR (purple). (c) Similar to panel b, but for ocean drifter observations with central gaps. The 1c case is shown without hatching, while the 2-component (2c) case is shown with cross-hatching. (d) PDFs of the MSE for a single configuration, $\bar{\Delta}$, obtained from C-DM and GPR for 1c and 3c cases, for a central gap of size $50\tau_\eta$ in Lagrangian turbulence reconstruction. (e) The MSE, $\langle \Delta (t) \rangle$ as a function of time within the gap, for Lagrangian turbulence reconstruction using C-DM and GPR for 1c and 3c cases, with a central gap of size $50\tau_\eta$. Here $t_g$ represents the relative time position from the left gap edge, as shown in panel a. (f) Similar to panel e, but for ocean drifter observations with a central gap of size $360\tau_0$. Error bars represent the minimum and maximum values obtained for different velocity components.}
\end{figure*}

On the other hand, the {\it backward} process is designed to reconstruct the signal within the gap, ensuring that both the original statistical properties and the correlation with the specific measured data realization are accurately reproduced. Once the learning process has converged, the neural networks model the conditional one-step backward transition probability, defined as $p_\theta(\CV^{(n-1)}_g|\CV^{(n)}_g,\CV_m)$, for each of the $N$ backward steps, $n=N,\dots, 1$. As a result, the generative refilling process inside the gap is obtained by starting with pure Gaussian noise at $n=N$, $p(\CV^{N}_g)=\mathcal{N}(\bm{0},\bm{I})$, and applying the neural network to model all backward steps down to $n=1$:
\begin{equation}
p_\theta(\CV_g^{(0:N)}|\CV_m)=p(\CV^{(N)}_g)\prod_{n=1}^{N}p_\theta(\CV^{(n-1)}_g|\CV^{(n)}_g,\CV_m).
\label{eq:backproc}
\end{equation}
In Fig.\ref{fig:setup}a1 we show an example of a tracer trajectory gradually generated along the backward process within the gap, $G$, while conditioned on the measure, $M$. A detailed description of the training protocol and the loss function can be found in Sec.\ref{subsec:C-DM} of the Methods. \\
\\
{\sc {\bf Gaussian process regression (GPR).}} 
To assess the performance of the C-DM, we define a baseline in terms of a multivariate Gaussian process (GP) \cite{williams2006gaussian}. A GP is a collection of random variables, any finite subset of which follows a joint Gaussian distribution. In our context, these random variables correspond to the signal values at sampled points $t_k$. Consequently, the joint distribution of the measurements $\CV_m$ and the signals within the gap $\CV_g$ is expressed as:
\begin{equation}
    \begin{bmatrix}
        \CV_m \\
        \CV_g
    \end{bmatrix}
    \sim\CN
    \left(
        \begin{bmatrix}
            \mu_m \\
            \mu_g
        \end{bmatrix},
        \begin{bmatrix}
            C_{mm} & C_{mg} \\
            C_{gm} & C_{gg}
        \end{bmatrix}
    \right),
\end{equation}
where $\mu_m=\langle\CV_m\rangle$ is the vector representing the mean of the signal at all time instants $t_m$ within the region $M$, and $\mu_g$ is similarly defined for $t_g$ in the gap $G$. The matrix $C_{mg}=\langle(V_m-\mu_m)(V_g-\mu_g)\rangle$ denotes the covariances between all pairs of measurement and gap points, with $C_{mm}$, $C_{gg}$, and $C_{gm}$ similarly representing the other covariance components. All entries can be estimated by averaging over the training data. To refill the gap in unseen test data, given the measurements $\CV_m$, we can use Bayes' rule and apply a standard regression process to estimate the posterior distribution of the signals within the gap as \cite{williams2006gaussian, von2014mathematical}:
\begin{align}
    p_{GPR}(\CV_g|\CV_m)\to\CV_g\sim\CN(&\mu_g+C_{gm}C_{mm}^{-1}(\CV_m-\mu_m), \nonumber \\
    &C_{gg}-C_{gm}C_{mm}^{-1}C_{mg}).
\end{align}\\
\\
{\sc{\bf Dataset: 3D tracers.}}
Lagrangian trajectories for pointlike particles (tracers) are extracted from high-resolution direct numerical simulations (DNS) of homogeneous isotropic turbulence (HIT) in a 3D incompressible velocity field $\bm{u}(\bm{x},t)$, governed by the Navier-Stokes equations (NSE) within a cubic periodic domain. The position and velocity of each particle, $(\bm{X}(t),\bm{V}(t))$, are determined by the advection equation driven by the underlying flow velocity:
\begin{equation}
\label{eq:eqm}
    \dot{\bm{X}}(t)=\bm{V}(t)=\bm{u}(\bm{X}(t),t).
\end{equation}
A total of $N_p=327,680$ trajectories are used to generate the training and test sets, divided 90\%/10\%, with each trajectory spanning a duration of $T\simeq1.3\tau_L\simeq200\tau_\eta$ and sampled at a time interval of $dt_s\simeq0.1\tau_\eta$, where $\tau_L$ and $\tau_\eta$ are the largest and smallest characteristic times of the underlying turbulent flow (see Sec.\ref{subsec:DNS} of the Methods for details on the DNS). Consequently, each trajectory is discretized into $K=2000$ time instants. \\
\\
{\sc \bf{Dataset: 2D ocean drifters.}}
To account for realistic geophysical scenarios, we used a dataset collected at regular hourly intervals from satellite-tracked surface drifting buoys (drifters) from NOAA-funded Global Drifter Program (GDP) \cite{elipot2022hourly}. Drifters are approximately Lagrangian \cite{Centurioni2018}, incorporating both spatial and temporal variability as they passively follow ocean currents (see Fig.\ref{fig:setup}c). They have been used in numerous previous studies to investigate a wide range of oceanic processes and assess numerical models \cite{centurioni2004observations, centurioni2008permanent, poulain2015direct, corrado2017general, zhang2018evolution, liu2019wind, zaron2019baroclinic, yu2019surface, arbic2022near}. We used version 2.01 of the dataset, which contains 19,396 individual surface drifter trajectories from October 1987 to October 2022, with approximately 197 million position and velocity estimates derived using the method described in \cite{elipot2016global}. In this work, we specifically consider only the zonal and meridional velocity components of all trajectories, without distinguishing between drogued and undrogued drifters, thus providing a dataset with diverse statistical properties to challenge the reconstruction tools. Training exclusively on drogued drifters leads to an unstable process without improving the validation results. Further distinguishing between the two configurations would require significantly more trajectories than are available in the current dataset and is beyond the scope of this study. To set up the reconstruction problem for drifters, we divided individual velocity time series into as many non-overlapping 60-day segments as possible. This resulted in 116,486 60-day segments, with each segment containing $K=1440$ points, corresponding to a shortest resolved time scale, $\tau_0 = 1h$, and a largest time scale of 1440 hours. 
After removing segments with spurious data points of high velocity and acceleration, 115,450 segments remained, which were then divided 90\%/10\% into training and test sets. Central gaps of sizes $36\tau_0$ and $360\tau_0$ are considered for reconstruction.\\
\\
{\sc \bf{Results.}} We will mainly discuss the case of a central missing gap (see Fig.\ref{fig:PointRecons}a1). The case of a gap at the end of a trajectory (see Fig.\ref{fig:PointRecons}a2), which involves the prediction of open-ended content with less contextual information, is discussed in detail in the Supplementary Material, as is the case of interpolation. We consider velocity as the quantity to be reconstructed and infer the spatial trajectory by successive integration. For both 3D HIT tracers and 2D drifters, we attempted to reconstruct either a single component or all three or two components simultaneously to exploit cross-correlations. Note that for 3D HIT tracers, statistical isotropy applies, whereas the 2D drifter problem is anisotropic.

\textit{Pointwise Reconstruction.} To evaluate the reconstruction accuracy at each instant within the gappy region, we calculate the mean squared error (MSE) between the reconstructed velocity field $\widetilde{V}_i$ and the true velocity field $V_i$ in the gap region $G$. This is given by
\begin{equation}
	\Delta(t)=[\tV_i(t)-V_i(t)]^2,
\end{equation}
where $t\in G$, $i$ is one of the components $x,y,z$ for 3D signals, and $i=e,n$ for eastward and northward velocities for drifters. We introduce angle brackets $\langle\cdot\rangle$ to denote averaging over all test configurations and an overbar $\bar{\cdot}$ to denote integration over $t$ in $G$. Thus we define the mean MSE as a function of $t$ within the gap, $\langle\Delta(t)\rangle$, and the mean MSE for a single trajectory as:
\begin{equation}
	\bar{\Delta}=\int_G\Delta(t)\,dt,
\end{equation}
with $\langle\bar{\Delta}\rangle$ representing the global MSE. All pointwise errors are normalized by a factor defined in terms of the total kinetic energies of the ground truth and the reconstructed signal:
\begin{equation}\label{equ:norm}
	Z=\langle\int_G(\tV_i)^2\,dt\rangle^{1/2}\langle\int_GV_i^2\,dt\rangle^{1/2},
\end{equation}
where for the 1-component (1c) case, different components are considered as separate configurations, while for the multi-component case, the energies in Eq.~(\ref{equ:norm}) are obtained from the average over all components $i$, resulting in the same $Z$ for both cases. In addition, $\Delta$ is calculated for data batches consisting of the same components in the test data, allowing us to generate error bars to quantify the variability of the reconstruction accuracy.

In Fig.\ref{fig:PointRecons}, we first present the global MSE obtained for the reconstruction of 3D Lagrangian tracers for different gap sizes $T_g$ (see panel a), ranging from window lengths comparable to the shortest turbulent time scales, $\sim \tau_\eta$, to windows as large as the longest turbulent correlation times, $\sim 100 \tau_\eta$ (panel b). Note that while the C-DM performs comparably to the linear GPR for the small gap size, we observe a small but systematic improvement by the C-DM as the gap size increases. The advantage of the C-DM is significantly enhanced when the reconstruction is applied to all three components simultaneously (cross-hatched histograms) for the $T_g = 50 \tau_\eta$ case. A similar improvement is observed when comparing the MSE of our C-DM and GPR for the 2D oceanic drifters (panel c). In panel d, we show the distribution of the instantaneous MSE within the central gap (panel a1) of size $50 \tau_\eta$. Overall, it is clear that the C-DM outperforms GPR, exhibiting a lower probability of committing large errors and a higher probability of being close to the ground truth. Moreover, the improvement is particularly notable for extreme worst-case scenarios (i.e., high reconstruction errors), where the far-right tail of the error distribution is consistently an order of magnitude smaller for C-DM compared to GPR in the 3-component (3c) case. 
In panels e and f, we show the MSE as a function of the time instant $t_g$ within the gap $0 \le t_g \le T_g$. It is clear that the C-DM systematically outperforms GPR, with a small improvement (around 10\%) for the 1c case and a significant improvement for the 3c case for 3D tracers.

\begin{figure*}[htbp]
	\includegraphics[width=\linewidth]{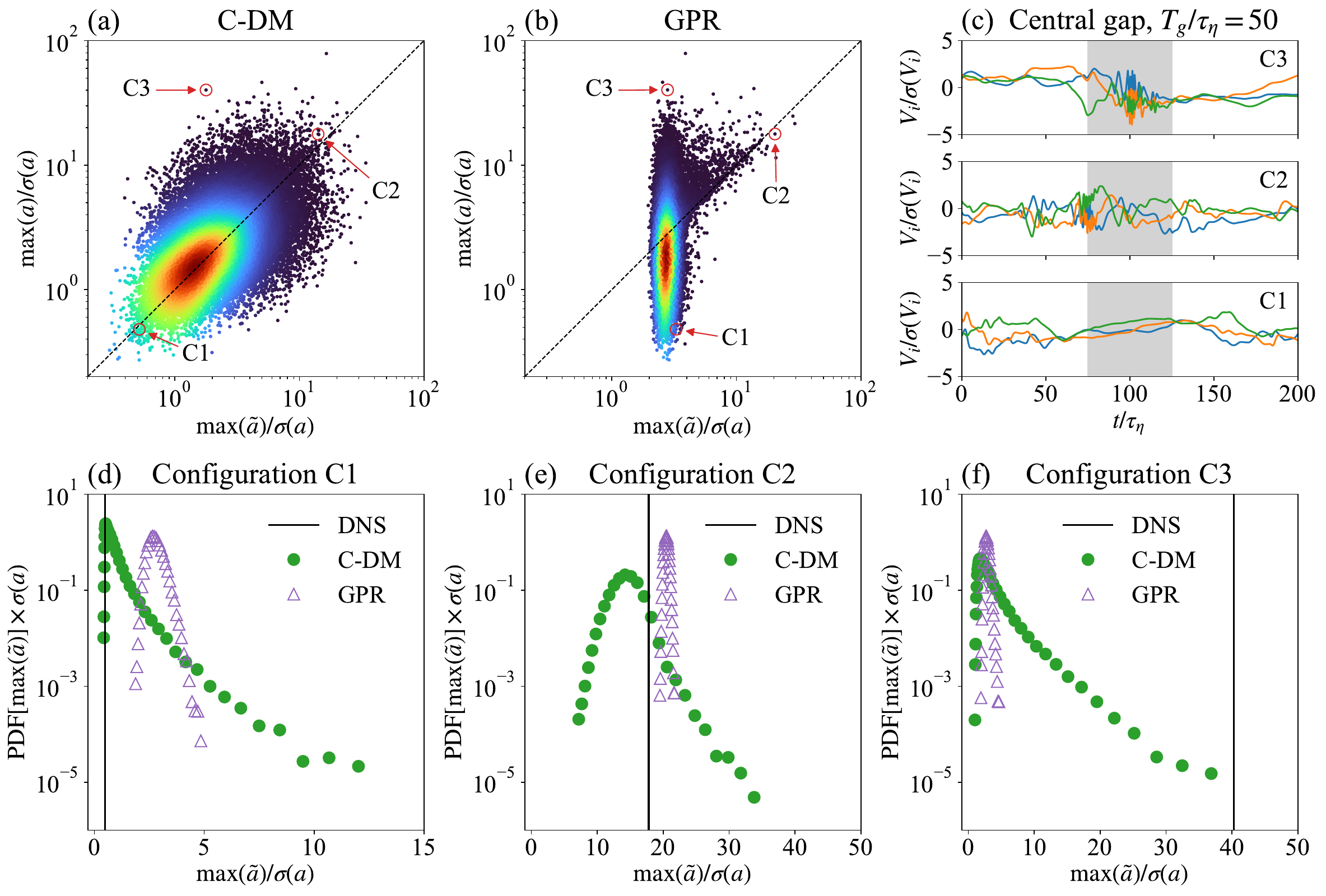}
     \caption{\label{fig:Max_acc} (a,b) Scatter plots of the maximum acceleration magnitude in a central gap of size $50\tau_\eta$, comparing the ground truth with reconstructions from (a) C-DM and (b) GPR. Colors represent the density of points in the scatter plot. Results are based on 32,768 test data, with one realization of the stochastic reconstructions for each configuration. Three specific configurations (C1, C2, C3), highlighted by red circles, are shown in (c). (d-f) PDFs of the maximum acceleration magnitude in the gap for the three fixed configurations: (d) C1, (e) C2, and (f) C3, from C-DM and GPR, with the ground truth DNS value marked by a vertical black line.}
\end{figure*}

We further assess the ability of different methods to reconstruct extreme events within the gap based on the given measurement configuration. Specifically, we focus on the largest values of the acceleration magnitude, $a=|\bm{a}|$, with $\tilde{a}$ representing the predicted values. These values are examined inside a central gap of size $50\tau_\eta$ for the Lagrangian turbulence case. The instantaneous particle acceleration is defined as
\begin{equation}
	a_i(t)=\dot V_i (t) ,
\end{equation}
which is known to possess extremely strong deviations from Gaussian statistics and fat tails, being connected to fluctuations at the highest turbulent frequency. In Fig.\ref{fig:Max_acc}a,b, we present scatter plots of the largest acceleration magnitudes from the original data and the predicted values from C-DM and GPR within the central gap. C-DM shows a strong correlation between the original and predicted values, while GPR exhibits little dependence of the predicted values on the original ones. Although the reconstruction methods are stochastic, these results are based on only one realization for each of the 32,768 test configurations. To evaluate the impact of stochasticity on the prediction performance of each method, we selected three specific configurations (Fig.\ref{fig:Max_acc}c) with increasing values of $\mathrm{max}(a)$ in the gap, marked by red circles in Fig.\ref{fig:Max_acc}a,b. For each fixed measurement, we generated 81,920 reconstructions, and in Fig.\ref{fig:Max_acc}d-f we plot the PDFs of the predicted $\mathrm{max}(\tilde{a})$ from both C-DM and GPR, with the ground truth DNS values shown as vertical black lines. 
For configuration C1, where the velocity variation within the gap is smoother and easier to reconstruct, C-DM has a distribution with a peak that better matches the DNS value (Fig.\ref{fig:Max_acc}d). For configuration C2, Fig.\ref{fig:Max_acc}e shows that GPR gives a large predicted $\mathrm{max}(\tilde{a})$ but produces a much narrower PDF, probably due to GPR's tendency to overshoot near the boundary, where extreme events occur at the left edge of the gap (as shown later in Fig.\ref{fig:ProbRecs_lagr}b and related discussion). In contrast, C-DM shows a wider PDF with a higher probability around the DNS value. 
For configuration C3, both methods fail due to the absence of a complete vortex structure inside the gap, with the measurements being too smooth and showing little correlation with the extreme events inside. However, Fig.\ref{fig:Max_acc}f shows that C-DM still has a chance to predict events of similar intensity within the gap.

\begin{figure*}[htbp]
	\includegraphics[width=\linewidth]{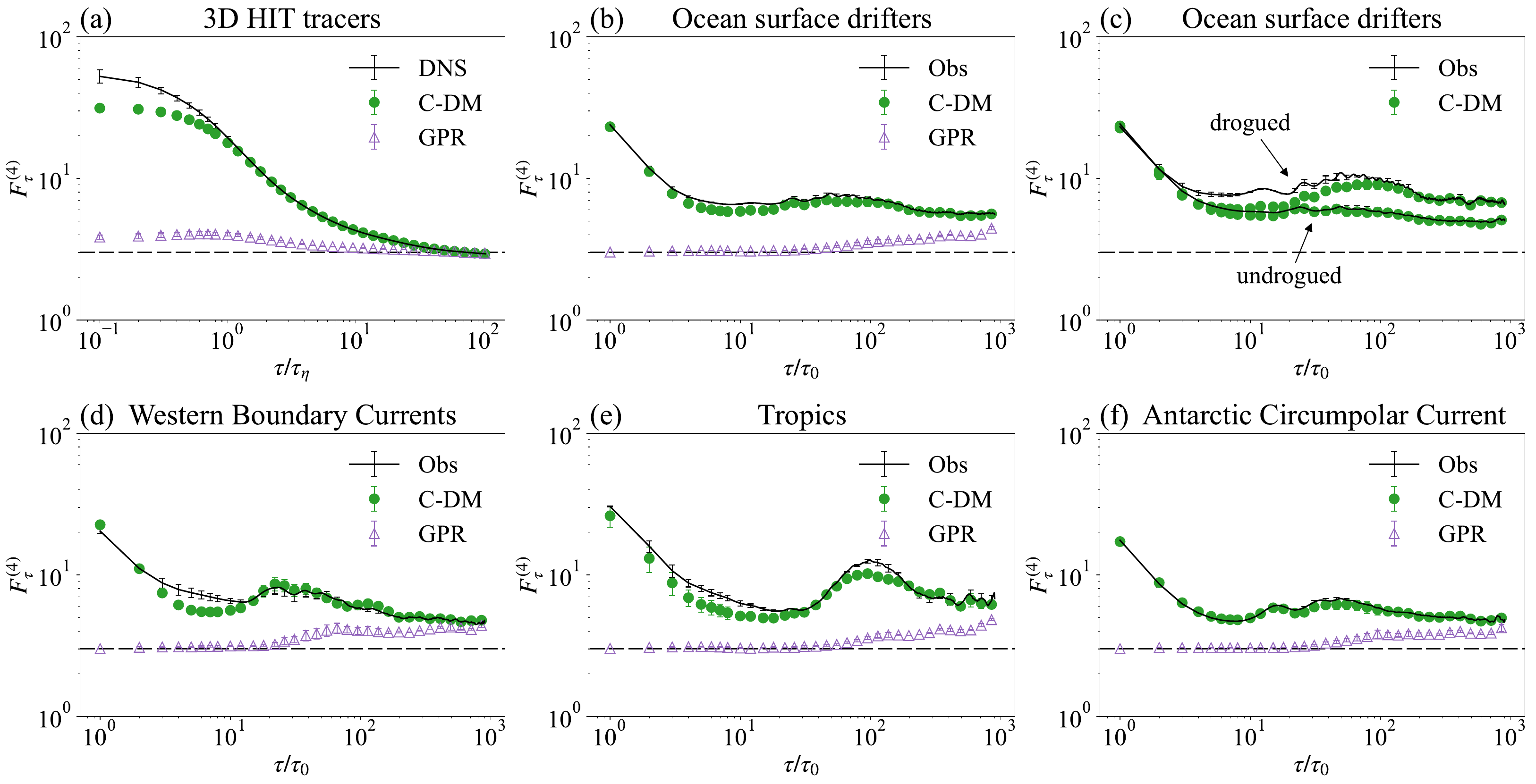}
	\caption{\label{fig:Flatness} (a) The fourth-order flatness, $F_\tau^{(4)}$, for 3D tracers from the ground truth DNS, C-DM and GPR reconstructions with a central gap of size $50\tau_\eta$. (b) $F_\tau^{(4)}$ for ocean drifter observations (Obs) with a central gap of size $360\tau_0$. (c) Same as panel b, but comparing Obs and C-DM reconstructions for fully drogued (top two) and undrogued (bottom two) drifters. (d-f) Regional $F_\tau^{(4)}$ conditioned on trajectories from the WBC (d), TRO (e) and ACC (f) regions, corresponding to regions A, B and C in Fig.\ref{fig:setup}c, respectively. Error bars are estimated from the spread between different velocity components.}
\end{figure*}

\begin{figure*}[htbp]
	\includegraphics[width=0.8\linewidth]{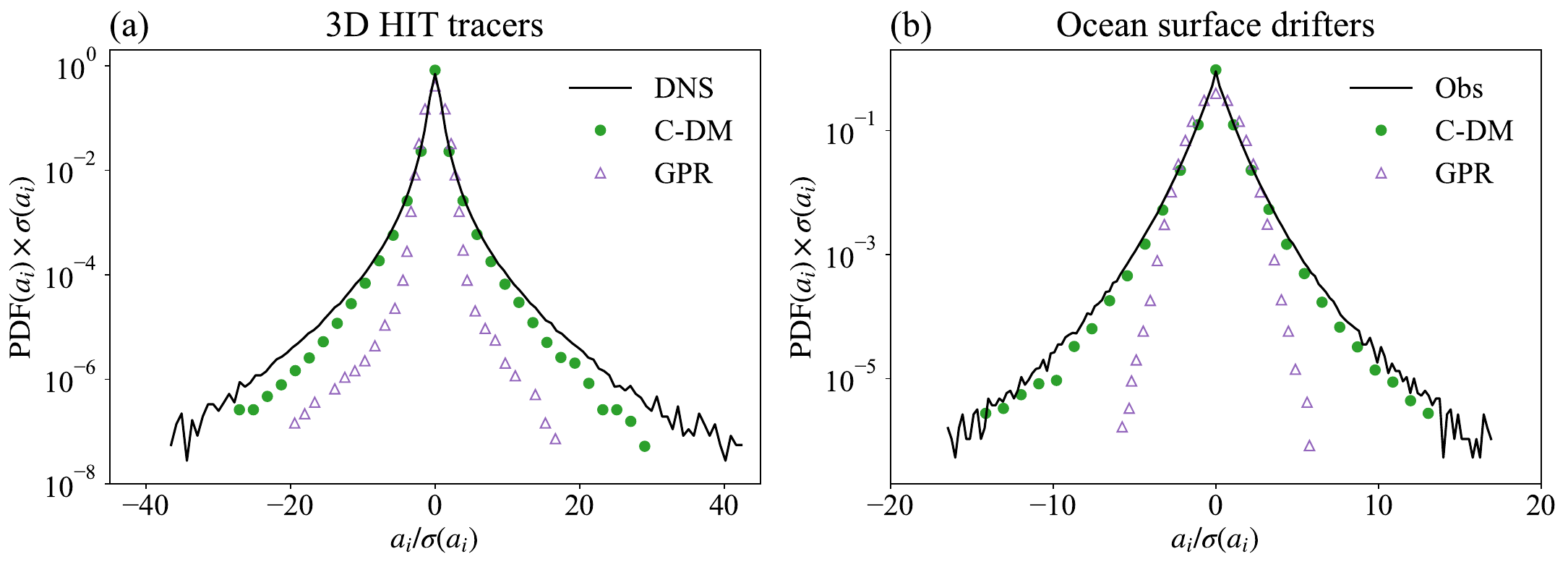}
	\caption{\label{fig:PDFs_a1c} (a) Standardized PDFs of a generic component of acceleration, $a_i$, for ground-truth DNS data (black line), C-DM reconstructed data (green solid circles) and GPR reconstructed data (purple hollow triangles) within a central gap of size $50\tau_\eta$ for Lagrangian turbulence reconstruction. (b) Similar to panel a, but for ocean drifter observations with a central gap of size $360\tau_0$.}
\end{figure*}

\begin{figure*}[htbp]
	\includegraphics[width=0.8\linewidth]{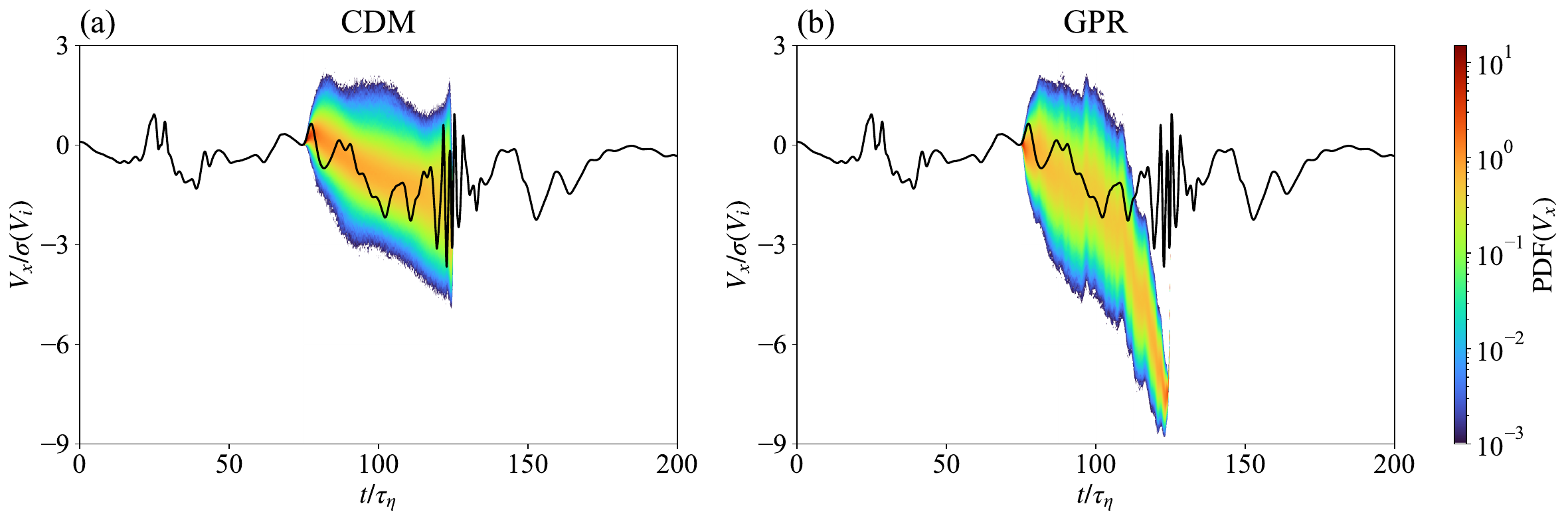}
	\caption{\label{fig:ProbRecs_lagr} Marginal PDFs of the $x$ velocity component from stochastic reconstructions for Lagrangian turbulence in a central gap region of size $50\tau_\eta$, focusing on a configuration where extreme events extend beyond the right edge of the gap. The reconstructions are derived from C-DM (a) and GPR (b), with the ground truth realization shown as a black line for reference.} 
\end{figure*}

\textit{Statistical Properties.} 
Given the wide range of time scales that characterise the signal, it is challenging to accurately reconstruct the signal in the $L_2$ sense well inside the gap, where correlations with the measurements are small. Therefore, a robust reconstruction method should aim to probabilistically reproduce the correct statistical properties, rather than focusing solely on pointwise accuracy. \\
The set of multiscale statistical properties used to evaluate the quality of the reconstruction is based on the velocity increment at different time lags $\tau$,
\begin{equation}
\label{eq:deltav}
	\delta_\tau V_i(t)=V_i(t+\tau)-V_i(t),
\end{equation}
conditioned to have at least one time instant inside the gap. 
From the instantaneous increment we can define the 
Lagrangian structure functions as
\begin{equation}
	S_\tau^{(p)}=\langle \overline{\delta_\tau V_i^p}\rangle.
\end{equation}
We can further calculate the generalized $p$-th order flatness as
\begin{equation}
	F^{(p)}_\tau=S^{(p)}_\tau/[S^{(2)}_\tau]^{p/2}.
\end{equation}
To illustrate, we present the results for the multi-component case with a central gap, where the gap size is $50 \tau_\eta$ for the 3D tracers and $360 \tau_0$ for the 2D drifters. \\
In Fig.\ref{fig:setup}b, we show the PDFs of the velocity increments in Eq.~(\ref{eq:deltav}) for different time lags $\tau$ for the 3D tracers. The accuracy of the C-DM in reproducing fluctuations of all intensities across all time lags is remarkable. Similarly, in panel d of the same figure, we show the PDF of the eastward single-point velocity for drifters in the three different geographic regions (A-C). Here again, the agreement between the ground truth observations and the C-DM generation is remarkable. \\
In Fig.\ref{fig:Flatness}, we present the 4th-order flatness for both datasets, comparing the ground truth statistics inside the gap with those reconstructed by our C-DM and GPR models. Panels a and b clearly show that C-DM captures data variability significantly better than GPR, with near-perfect agreement with DNS for the 3D dataset and observations for the 2D dataset at the global level. For the oceanic drifters, panel c distinguishes between drogued and undrogued cases, while panels d-f show results conditioned on the three different regions (A-C) highlighted in Fig.\ref{fig:setup}c. The C-DM reconstructed 4th-order flatness for undrogued drifters aligns better with observations (Fig.\ref{fig:Flatness}c), probably due to the dominance of undrogued drifters in the training dataset ($60\%$ of trajectories are fully undrogued, while $30\%$ are fully drogued). In addition, the flatness for drogued data shows more intermittent behaviour (i.e. further from the value of 3 given by Gaussian statistics over scales), making it more difficult to learn. The three plots in Fig.\ref{fig:Flatness}c-e show that the C-DM is able to capture the strong regional variability of the statistical properties. For the Western Boundary Currents and the Tropics, the C-DM reconstructed flatness shows excellent agreement with the observations for time scales larger than the main tidal periods around 12--24 hours (panels d and e). Small differences are observed in the Tropics, particularly in the near-inertial band between 40 and 200 hours (panel e). Remarkably good agreement between observations and C-DM reconstructions is also found in the Antarctic Circumpolar Current (panel f). Notice the clear failure of GPR, which, by definition, is able to generate only signals with a Gaussian self-similar refilling, conditioned to the measured data. \\
In Fig. \ref{fig:PDFs_a1c}, we present one of the most stringent statistical tests to evaluate the expressivity of the stochastic model by comparing the PDFs for the acceleration of both 3D and 2D signals (panels a and b, respectively). Once again, the ability of the C-DM to reproduce extreme events is remarkable, capturing values up to 40 and 20 times the standard deviation for the two data sets. In contrast, the GPR shows significantly weaker performance.

\begin{figure*}[htbp]
	\includegraphics[width=\linewidth]{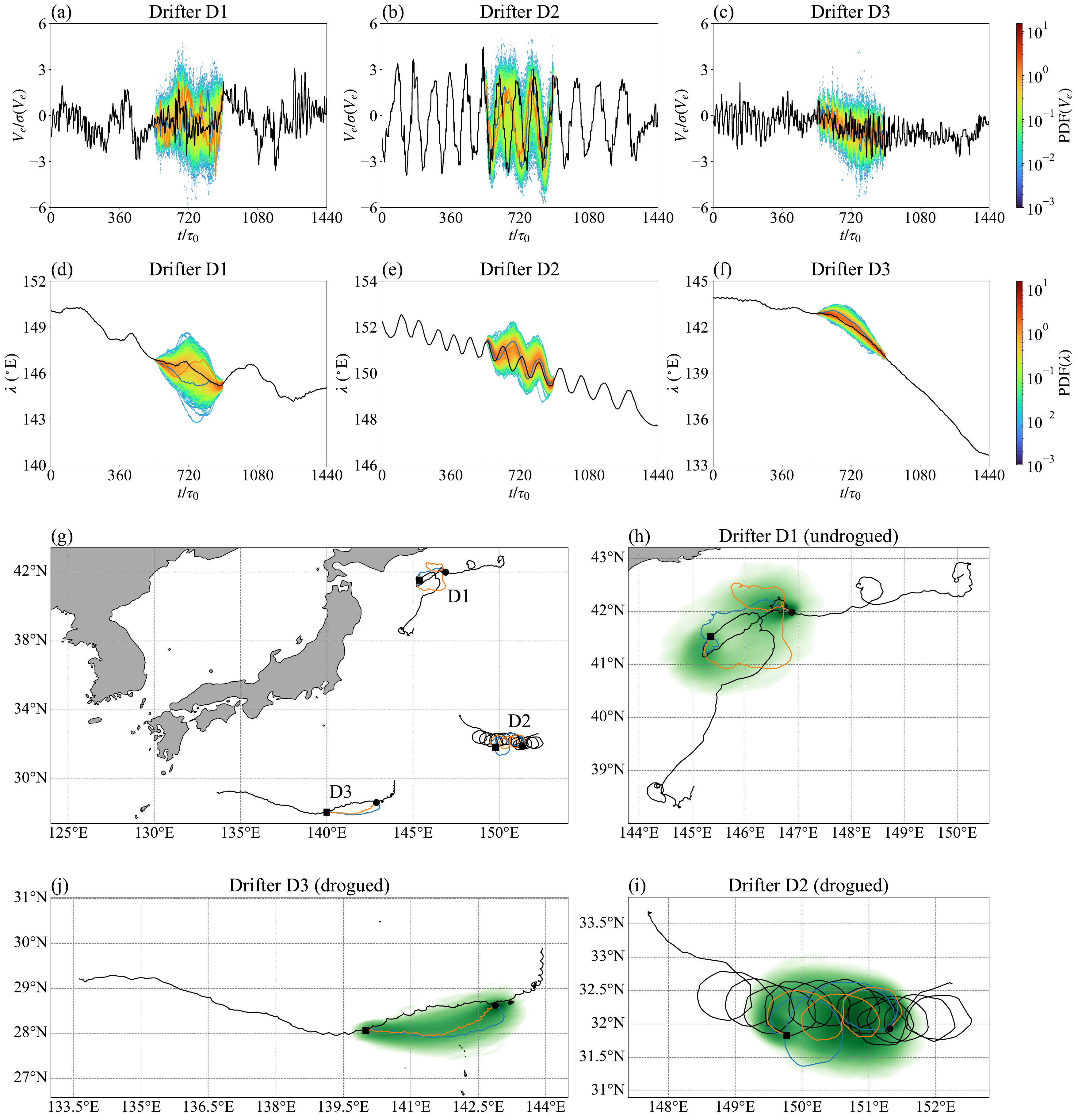}
	\caption{\label{fig:ProbRecs} (a-c) Marginal PDFs of the eastward velocity component from stochastic reconstructions using C-DM for three oceanic drifters (shown in panel g) in a central gap region of size $360\tau_0$. (d-f) Corresponding marginal PDFs of longitude $\lambda$, where positions (longitude and latitude) are derived by integrating the reconstructed velocity signals on the sphere. (g) Three partially observed trajectories in the Kuroshio Current (black lines) with a central gap of $360\tau_0$, marked by a black circle (start point) and a black square (end point). (h-j) Zoomed-in views of each trajectory, also showing the density of reconstructed trajectory points in green clouds. In all panels, 1,024 trajectories are selected out of 81,920 reconstructions, with the closest position to the ground truth at the right end of the gap. The ground truth is shown as a black line, while the two closest reconstructions from the C-DM model are displayed as blue and orange lines.}
\end{figure*}

\textit{Uncertainty quantification.} 
The stochastic properties of the C-DM naturally allow for {\it uncertainty quantification} by generating many different signal instances within the gap region $G$, for a given set of measurements in $M$. In Fig.\ref{fig:ProbRecs_lagr}, we present the distribution of velocity profiles for the 3D tracer case, focusing on the $x$-velocity component of a trajectory, selected for its strong vortical event near the final end of the gap, characterized by extreme non-Gaussian fluctuations across the gap boundary. 
The comparison between panel a, obtained with C-DM, and panel b, obtained with GPR, shows the improved ability of C-DM to capture the correct fluctuations within the gap, in contrast to the strong overshooting exhibited by the GPR cloud near the extreme event. This clearly shows the limitations of the Gaussian assumption.
In Fig.\ref{fig:ProbRecs}, we present statistical refilling results using C-DM for three oceanic drifters (D1, D2, D3) in the Kuroshio Current (see panels g-j for geographical locations and drogue status). By integrating velocity signals on the sphere, we reconstruct the positions (longitude and latitude). Out of 81,920 reconstructions generated by C-DM, 1,024 were selected based on their proximity to the ground truth at the end of the gap (black squares in panels g-i). Panels a-c show the marginal PDFs at different time instants of the eastward velocity. The ability of the C-DM to accurately capture `fluctuations' is qualitatively evident, as it adapts to the varying background measurements and refills the signal with frequencies consistent with the observed data. In panels d-f, we show the marginal PDFs of the reconstructed longitude, $\lambda$, demonstrating C-DM's ability to reconstruct also spatial coordinates from the inferreed velocity signals. Panels h-j show the density of trajectory points, with the green cloud representing the spread of these points across the reconstructed paths. 
In all panels of Fig.\ref{fig:ProbRecs}, the ground truth is shown as black lines, while the two best reconstructions (closest to the ground truth at the end of the gap) are shown as blue and orange lines. It is worth noting that for drifter D2, which tracks currents rotating at near-inertial frequency due to a likely sudden shift in wind stress direction and intensity, the C-DM method reconstructs the trajectory reasonably well, probably because the oscillations are present both before and after the gap (panels b, e and i). 
Similarly, for drifter D1, the peak of the trajectory distribution seems to closely follow the ground truth, as if the model captures the subtle undulation in the longitudinal coordinate along the entire trajectory (panel d). Finally, for drifter D3, the motion along the longitudinal direction is much more linear, and the reconstructed trajectories easily match this linearity (panel f). A more regionally focused segregation of the training dataset could potentially improve the accuracy of the results. This suggests how the stochastic nature of the velocity refilling can be leveraged to obtain realistic missing spatial signals for trajectories where the positions at the beginning and end of the gap are known.
\\
\\
{\sc \bf{Conclusions.}} 
A novel application of conditional diffusion models for stochastic reconstruction of trajectories along 3D turbulent tracers and 2D oceanic drifters from NOAA-funded Global Drifter Program is proposed. Superiority over quantitative benchmarks obtained using Gaussian process regression is demonstrated in terms of both pointwise reconstruction using MSE and statistical expressivity. The latter is demonstrated by assessing highly non-Gaussian multiscale properties, as measured by the flatness of velocity increment distributions over a wide range of time scales spanning more than three decades, as well as by the PDF of acceleration. For 3D tracers, the stochastic C-DM is able to correctly capture acceleration fluctuations up to 40 times the standard deviation, i.e. including extreme events. 
Our model is proven to be robust enough to capture varying statistical properties across different geographical regions for oceanic drifters and can be used to generate a set of {\it optimal} paths to estimate the drifter trajectories during the `blind' measurement window, suggesting promising applications for data augmentation of geophysical ocean surface measurements. However, generalization to cases where unknown observables strongly affect the local trajectory could significantly impact inference accuracy, especially if not augmented by regional information. This is particularly relevant in scenarios such as strong wind bursts occurring within the gap. The model is flexible enough to be applied to a variety of different gap geometries and locations and in many different fields, including other Lagrangian turbulence problems such as 2-particle and multi-particle dispersions, charged particles in astrophysical applications, active matter (e.g. pedestrian dynamics), and whenever data need to be repaired or denoised. The method has also been generalised for 2D Eulerian turbulence data augmentation \cite{li2023diffusion}.
Open key problems remain, related to the scaling properties of the architecture with respect to the complexity and amount of training data \cite{peebles2023scalable}, as well as the issue of model collapse when unconditioned or conditioned data augmentation is used to train new generations of models \cite{shumailov2024ai}. Comparisons with other data-driven approaches, such as gappy POD, extended POD, and generative adversarial network (GAN) \cite{everson1995karhunen, boree2003extended, goodfellow2014generative, li2023multi, guastoni2021convolutional, cuellar2024three}, as well as model-based methods \cite{friedrich2020stochastic, lubke2023stochastic} are possible. However, a systematic ranking of all methods is beyond the scope of this work. Such an evaluation would require a community-wide effort to establish benchmarks and grand challenges -- an effort that is still lacking for realistic problems involving the inference of highly chaotic and turbulent systems such as those studied here.\\
\\
{\sc \bf{Methods}}
\subsection{Conditional DMs for reconstruction}\label{subsec:C-DM}
\begin{figure*}[htbp]
    \begin{tikzpicture}
        \node[inner sep=0] (imgA) {
        \includegraphics[width=1.0\textwidth]{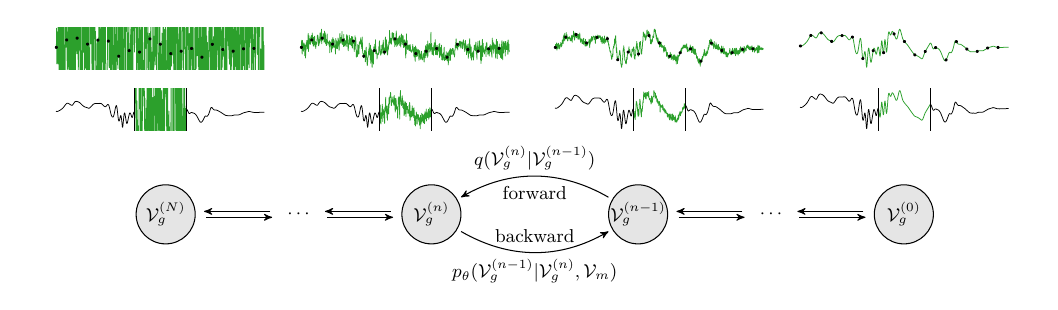}};
        \node[anchor=north west] at ([yshift=0.8em]imgA.north west) {\fontsize{10}{12}\selectfont (a)};
        \node[inner sep=0, below=0cm of imgA] (imgB) {
        \includegraphics[width=1.0\textwidth]{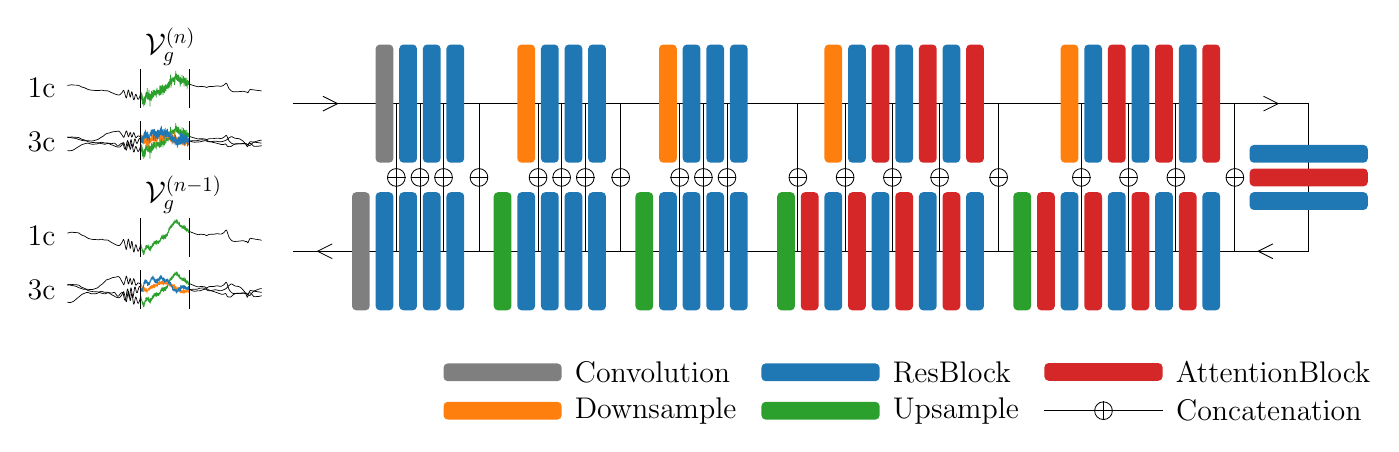}};
        \node[anchor=north west] at ([yshift=0.8em]imgB.north west) {\fontsize{10}{12}\selectfont (b)};
    \end{tikzpicture}
    \caption{\label{fig:CDM} (a) Schematic of the C-DM protocol for turbulent signal reconstruction. In the forward process (from right to left), noise is gradually added to the signal within the unknown region, $\CV_g=\CV_g^{(0)}$, over $N$ steps according to a predefined schedule. The noisy signal at step $n$ within the gap, $\CV_g^{(n)}$, is represented by green lines. Partial measurements, $\CV_s$, are represented by black points (for interpolation, top) or black lines (for gap filling, middle). In the backward process (from left to right), reconstruction starts with pure noise within the gap, $\CV_g^{(N)}$, which is combined with the measurements to progressively denoise the missing information using the trained neural network. (b) The U-Net architecture of the neural network for a denoising step, $p_\theta(\CV^{(n-1)}_g|\CV^{(n)}_g,\CV_m)$. The noisy signal at step $n$, $\CV_g^{(n)}$, is first combined with the measurements to form a complete signal, which is then concatenated along the channel dimension with a signal consisting of the measurements outside the gap and random noise inside the gap. The network output has the length of a full signal, and only the part inside the gap is filtered out as $\CV_g^{(n-1)}$.}
\end{figure*}

Here we give a detailed description of the C-DMs used in this work to reconstruct Lagrangian trajectories from partial velocity measurements. As briefly introduced above, C-DMs consist of two main processes: the forward and the backward process, see Fig.\ref{fig:CDM}a. 
The one-step forward transition probability can be written as:
\begin{equation}
    q(\CV^{(n)}_g|\CV^{(n-1)}_g)\to\CV^{(n)}_g\sim\mathcal{N}(\sqrt{1-\beta_n}\CV^{(n-1)}_g,\beta_n\bm{I}),
\end{equation}
where the initial realization inside the gap coincides with the ground truth signal, $\CV^{(0)}_g =\CV_g $, and the variance schedule, $\{\beta_1,\ldots,\beta_N\}$, is predefined to progressively destroy the correlations between the data in the gap and the measured signal, $\CV_m $, resulting in a smooth transition to the pure Gaussian state, $\CV^{(N)}_g \sim\mathcal{N}(0,\bm{I})$. 
We can formally express the forward process as
\begin{equation}
q(\CV^{(1:N)}_g|\CV^{(0)}_g)\coloneqq\prod_{n=1}^{N}q(\CV^{(n)}_g|\CV^{(n-1)}_g),
\end{equation}
where the notation $\CV^{(1:N)}_g$ denotes the entire sequence of noisy trajectories, $\{\CV^{(1)}_g,\CV^{(2)}_g,\dots,\CV^{(N)}_g\}$, generated from the initial trajectory $\CV^{(0)}_g$ in the gap.
Note that the data within the measurement region is never accessed in the forward process. \\
The backward process models each step of the denoising conditional probability given measurements outside the gap, $p_\theta(\CV^{(n-1)}_g|\CV^{(n)}_g,\CV_m)$, using a neural network with parameters $\theta$. Once trained, the C-DM reconstructs the trajectory within the gap, starting from pure Gaussian noise, $\CV^{(N)}_g$, and conditioning on the measurements, $\CV_m$, by iteratively reversing the forward diffusion process as introduced in Eq.~(\ref{eq:backproc}). \\
In the continuous diffusion limit, where a large number of diffusion steps are used and the noise variance $\beta_n$ is chosen to be small, we can assume that the backward transition probability, $p_\theta(\CV^{(n-1)}_g|\CV^{(n)}_g, \CV_m)$, follows the same Gaussian functional form as the forward step~\cite{feller2015theory, sohl2015deep}. The neural network is then tasked with predicting the mean, $\mu_\theta(\CV^{(n)}_g, \CV_m, n)$, and the covariance, $\Sigma_\theta(\CV^{(n)}_g, \CV_m, n)$, for each denoising step. Following~\cite{ho2020denoising}, we set $\Sigma_\theta = \beta_n \mathbf{I}$, using step-dependent constants that remain untrained. Consequently, each one-step backward sampling is reformulated as:
\begin{equation}
\label{eq:nn}
    p_\theta(\CV^{(n-1)}_g|\CV^{(n)}_g) \to \CV^{(n-1)}_g \sim \mathcal{N}(\mu_\theta(\CV^{(n)}_g, \CV_m, n), \beta_n \bm{I}).
\end{equation}
The model optimization is performed by minimizing a variational upper bound on the negative log-likelihood, as in standard generative DMs. The additional conditioning is explicitly expressed in both the target distribution, $p(\CV_g|\CV_m)$, and the approximated distribution, $p_\theta(\CV^{(0)}_g|\CV_m)$:
\begin{equation}
\label{equ:nll}
\mathbb{E}_{p(\CV_g|\CV_m)}[-\log(p_\theta(\CV^{(0)}_g|\CV_m))].
\end{equation}
A detailed derivation of the loss function can be found in~\cite{li2023diffusion, li2024synthetic}.

The backbone neural network for the C-DMs in this work is based on a U-Net architecture \cite{ronneberger2015u}, building upon the design previously used for unconditional Lagrangian turbulence generation \cite{li2024synthetic}. To incorporate conditioning on the measurements, the input is modified as a combination of the measurement data and the noisy generation inside the gap, $\CV_m \cup \CV_g^{(n)}$, with additional channels that consist of the measurement and random noise within the gap. Fig.~\ref{fig:CDM} provides a graphical representation of the U-Net architecture and its role in the C-DM refilling process. 
The U-Net architecture consists of two main components: a downsampling stack and an upsampling stack, which are arranged symmetrically. Both stacks perform four steps of downsampling and upsampling respectively, resulting in five stages from left to right for each stack. Across these five stages, the residual blocks are configured with channel sizes of $\{1C, 1C, 2C, 3C, 4C\}$, where $C$ is 128. The last two stages of both stacks contain multi-head attention blocks, each with four heads. Connecting the downsampling and upsampling stacks is an intermediate module containing two residual blocks surrounding a central four-head attention block. The optimal noise schedule from \cite{li2024synthetic} is applied, with a total of $N = 800$ diffusion steps. Each specific C-DM case is trained with a batch size of 256 on four NVIDIA A100 GPUs, taking approximately 24 hours.

\subsection{Navier–Stokes simulations for Lagrangian tracers}\label{subsec:DNS} 
To evolve the turbulent flow advecting the Lagrangian tracers, we numerically solve the 3D incompressible NSE:
\begin{equation}
	\begin{cases}
		\partial_t\bm{u}+\bm{u}\cdot\nabla\bm{u}=-\nabla p+\nu\Delta\bm{u}+\bm{F} \\
		\nabla\cdot\bm{u}=0
	\end{cases},
\end{equation}
where $\bm{u}$ is the Eulerian velocity field, $p$ is the pressure, and $\nu$ is the fluid viscosity \cite{frisch1995turbulence}. We used a standard pseudo-spectral solver, fully dealiased with the two-thirds rule. The flow is driven by homogeneous and isotropic forcing, $\mathbf{F}$, applied at large scales via a second-order Ornstein-Uhlenbeck process \cite{forcingsawford} to reach a statistically steady state, after which particles are introduced into the system. Further details on the simulation can be found in \cite{biferale2023turb}. Lagrangian tracer integration is performed using a 6th-order B-spline interpolation to obtain fluid velocity at the particle positions, combined with a 2nd-order Adams-Bashforth time-marching scheme \cite{van2012efficiency}. Lagrangian trajectories are recorded at intervals of $dt_s = 15 dt \simeq 0.1 \tau_\eta$ \cite{calascibetta2023optimal}. Table~\ref{table:Table1} summarizes the simulation parameters.

\begin{table}[h!]
	\centering
	\begin{tabular}{|c|c|c|c|}
		\hline
		$N_L$ & $L$    & $dt$               & $\nu$ \\  
		1024  & $2\pi$ & $1.5\times10^{-4}$ & $8\times10^{-4}$ \\ \hline
		$\epsilon$  & $\tau_\eta$                & $\eta$                     & $R_\lambda$ \\
		$1.8\pm0.1$ & $(2.1\pm0.2)\times10^{-2}$ & $(4.2\pm0.1)\times10^{-3}$ & $\simeq 310$ \\ \hline
		${ N}_{p}$ & $dt_{s}$            & $T$   & $K$ \\
		$327680$   & $2.25\times10^{-3}$ & $4.5$ & $2000$ \\ \hline
	\end{tabular}
	\caption{\textbf{Eulerian parameters:} $N_L$ is the number of grid points in each spatial dimension. $L$ is the physical size of the cubic box. $dt$ is the time step used in the DNS integration. $\nu$ is the kinematic viscosity. $\epsilon=\nu\langle\partial_i u_j\partial_i u_j\rangle$ represents the mean energy dissipation, averaged over time and space. $\tau_\eta=\sqrt{\nu/\epsilon}$ is the Kolmogorov dissipative time. $\eta=(\nu^3/\epsilon)^{1/4}$ is the Kolmogorov dissipative scale. $R_\lambda=u_{rms}\lambda/\nu$ is the Taylor-scale Reynolds number, where $u_{rms}$ is the root mean squared velocity, and $\lambda=\sqrt{5E_{tot}/\Omega}\simeq0.14$ is the Taylor-scale. Here, $E_{tot}\simeq4.5$ and $\Omega\simeq1200$ represent the mean energy and enstrophy, respectively. $\tau_L=L/u_{rms}\simeq 3.5$ is the integral time scale. \textbf{Lagrangian parameters:} $N_{p}$ is the total number of trajectories. $dt_{s}$ is the time interval between two consecutive Lagrangian dumps. $T$ is the total duration of each trajectory, and $K=T/dt_{s}$ is the total number of points per trajectory.}
	\label{table:Table1}
\end{table}

{\sc \bf{Data availability.}} The 3D HIT tracer trajectories used in this study are available for download from the open access Smart-TURB portal (\url{http://smart-turb.roma2.infn.it}) under the TURB-Lagr repository \cite{biferale2023turb}. Additionally, both these trajectories and the processed segments of velocities for oceanic drifters, as well as the initial positions of these segments, are available on the INFN Open Access Repository (\url{https://doi.org/10.15161/oar.it/211740}) \cite{tianyi_li_2024_211740}. \\
\\
{\sc \bf{Code availability.}} The code to train the C-DM model and perform the reconstruction can be found at \url{https://github.com/SmartTURB/C-DM-lagr}. We will provide reviewers with access to the code repository during the peer review process. The repository will be made public once the paper is published.\\ 
\\
{\sc \bf{ Acknowledgments.}}
We thank M. Sbragaglia for collaboration in a early stage of this work. We are also grateful to Jeremiah L\"ubke for valuable discussions. This work was supported by the European Research Council (ERC) under the European Union’s Horizon 2020 research and innovation programme Smart-TURB (Grant Agreement No. 882340). {L.C. was supported by NOAA grant NA20OAR4320278 ``The Global Drifter Program''.}

\bibliography{apssamp}

\end{document}